\documentclass[aps,prl,twocolumn,groupedaddress,amsmath,amssymb]{revtex4-1}
\usepackage{graphicx}  
\usepackage{dcolumn}   
\usepackage{bm}        
\usepackage{verbatim}   
\usepackage{hyperref}

\begin{document}

%
\title{Optoacoustic inversion via Volterra kernel reconstruction}
%
\author{O.\, Melchert}
\email{oliver.melchert@hot.uni-hannover.de}
\author{M.\, Wollweber}
\author{B.\, Roth}
%
\affiliation{
  Hannover Centre for Optical Technologies (HOT), 
  Leibniz Universit\"at Hannover, 
  D-30167 Hannover, Germany
}
%

\date{\today}

\begin{abstract}
In this letter we address the numeric inversion of optoacoustic signals to
initial stress profiles. Therefore we put under scrutiny the optoacoustic
kernel reconstruction problem in the paraxial approximation of the underlying
wave-equation.  We apply a Fourier-series expansion of  the optoacoustic
Volterra kernel and obtain the respective expansion coefficients for a given
``apparative'' setup by performing a gauge procedure using synthetic input
data.  The resulting effective kernel is subsequently used to solve the
optoacoustic source reconstruction problem for general signals.  We verify the
validity of the proposed inversion protocol for synthetic signals and explore
the feasibility of our approach to also account for the diffraction
transformation of signals beyond the paraxial approximation.
\end{abstract}

\maketitle

The {\emph{inverse}} optoacoustic (OA) problem is concerned with the
reconstruction of ``internal'' OA properties from ``external'' measurements of
acoustic pressure signals.  In contrast to the {\emph{direct}} OA problem,
referring to the calculation of a diffraction-transformed pressure signal at a
desired field point for a given initial stress profile
\cite{Diebold:1990,*Diebold:1991,*Calasso:2001,Gusev:1993,Landau:1981,Colton:2013},
one can distinguish two inverse OA problems: (I.1) the {\emph{source
reconstruction problem}}, where the aim is to invert measured OA signals to
initial stress profiles upon knowledge of the mathematical model that mediates
the underlying diffraction transformation
\cite{Wang:2009,Colton:2013,Kuchment:2008}, and, (I.2) the {\emph{kernel
reconstruction problem}}, where the task is to reconstruct a proper OA
stress-wave propagator to account for the apparent diffraction transformation
shown by the OA signal.  While, owing to its immediate relevance for medical
applications
\cite{Xu:2006,*Wang:2012,*Yang:2012,*Wang:2013,*Wang:2014,*Stoffels:2015,*Stoffels:2015ERR},
current progress in the field of inverse optoacoustics is spearheaded by OA
tomography and imaging applications in line with (I.1)
\cite{Agranovsky:2007,*DeanBen:2012,*Belchami:2016,Norton:1981,*Xu:2005,*Burgholzer:2007},
problem (I.2) has not yet received much attention (note that quite similar
kernel reconstruction problems are well studied in the context of
inverse-scattering problems in quantum mechanics
\cite{Chadan:1989,*Apagyi:1997,*Munchow:1980,*Melchert:2006}).  However, under
ill-conditioned circumstances that prohibit a consistent description of the
stress-wave propagation or when the multitude of signals that form the
inversion input to common backpropagation approaches (see, e.g.,
Refs.\,\cite{Norton:1981,*Xu:2005,*Burgholzer:2007}) are simply inaccessible,
kernel reconstruction in terms of (I.2) provides an opportunity to yield a
reliable OA inversion protocol in terms of single-shot measurements. 

As a remedy, we here describe a numerical approach to problem (I.2), appealing
from a point of view of computational theoretical physics.  More precisely, in
the presented letter, we focus on the kernel reconstruction problem in the
paraxial approximation to the optoacoustic wave-equation, where we suggest a
Fourier-expansion approach to construct an approximate stress wave propagator.
We show that once (I.2) is solved for a given ``apparative'' setup, this then
allows to subsequently solve (I.1) for different signals obtained using an
identical apparative setup.  A central and reasonable assumption of our
approach is that the influence of the stress wave propagator on the shape
change of the OA signal is negligible above a certain cut-off distance.
After developing and testing the numerical procedure in the paraxial
approximation, we assess how well the inversion protocol carries over to more
prevalent optoacoustic problem instances, featuring the reconstruction for: (i)
the full OA wave-equation, (ii) non Gaussian irradiation source profiles, and,
(iii) measured signals exhibiting noise.

\paragraph{The direct OA problem.}
The dominant microscopic mechanism contributing to the generation of acoustic
stress waves is expansion due to photothermal heating \cite{Tam:1986}. In the
remainder we assume a pulsed photothermal source with pulse duration short
enough to ensure thermal and stress confinement \cite{Wang:2009}.  Then, in
case of a purely absorbing material exposed to a irradiation source profile
with beam axis along the $z$-direction of an associated coordinate system, a
Gaussian profile in the transverse coordinates $\vec{r}_\perp$ and nonzero
depth dependent absorption coefficient $\mu_{a}(z)$, limited to $z\geq 0$ and
varying only along the $z$-direction, the initial acoustic stress response 
to photothermal heating takes the form 
\begin{equation}
p_{\rm 0}(\vec{r})= 
        f_0\, \mu_{\rm a}(z) \exp\Big\{-|\vec{r}_\perp|^2/a_{\rm B}^2-
        \int_0^z \!\mu_{\rm a}(z^{\prime})~\mathrm{d}z^{\prime}\Big\}.
        \label{eq:p0}
\end{equation}
Therein $f_0$ and $a_{\rm B}$ signify the intensity of the irradiation source
along the beam axis and the $1/e$-width of the beam profile orthogonal to the
beam axis, respectively.
Given the above initial instantaneous acoustic stress field $p_{\rm
0}(\vec{r})$, the scalar excess pressure field $p(\vec{r},t)$ at time $t$ and
field point $\vec{r}$ can be obtained by solving the inhomogeneous OA wave
equation \cite{Gusev:1993,Wang:2009} 
\begin{equation}
\big[ \partial_t^2 - c^{2} \Delta \big]~p(\vec{r},t) = 
        p_0(\vec{r})~\partial_t\,\delta(t), 
        \label{eq:OAWaveEq}
\end{equation}
with $c$ denoting the sonic speed within the medium.
The acoustic near and far-field might be distinguished by means of the
diffraction parameter $D=2|z_{\rm D}|/(\mu_{\rm a} a_{\rm B}^2)$, 
where near and far-field are characterized by $D<1$ and $D>1$, respectively.

In the paraxial approximation where the full wave equation reduces to the
parabolic diffraction equation \mbox{$[\partial_\tau \partial_z -
(c/2)\Delta_\perp ]\,p=0$} \cite{Gusev:1993,Karabutov:1996}, it can be shown
that the time-retarded ($\tau=t+z_{\rm D}/c$) OA signal at a field point 
along the beam axis $p_{\rm D}(\tau)\equiv p(\vec{r}_{\rm D},t)$ 
can be related to the initial ($t=0$) on-axis stress profile 
$p_{\rm 0}(\tau)\equiv p_{\rm 0}(\vec{r}_\perp\!=\!0,z)$ via a Volterra
integral equation of $2$nd kind, reading \cite{Karabutov:1996}
\begin{align}
p_{\rm D}(\tau) = p_{\rm 0}(\tau) - \int_{-\infty}^{\tau}
        \!\mathsf{K}(\tau-\tau^\prime)\,p_{\rm 0}(\tau^\prime)
        \,\mathrm{d}\tau^\prime.
        \label{eq:OAVolterraInt}
\end{align}
Therein the Volterra operator features a convolution kernel
\mbox{$\mathsf{K}(\tau-\tau^\prime)=\omega_{\rm D} \exp\{-\omega_{\rm
D}(\tau-\tau^\prime)\}$},  mediating the diffraction transformation of the
propagating stress waves. The characteristic OA frequency $\omega_{\rm D}= 2 c
|z_{\rm D}|/a_{\rm B}^2$ effectively combines the defining parameters of the
apparative setup ${\bf p}_{\rm sys}\equiv (c,a_{\rm B},z_{\rm D})$.
Subsequently we focus on OA signal detection in backward mode, i.e.\ $z_{\rm D}
< 0$.

\begin{figure}[t!]
\begin{center}
\includegraphics[width=1.0\linewidth]{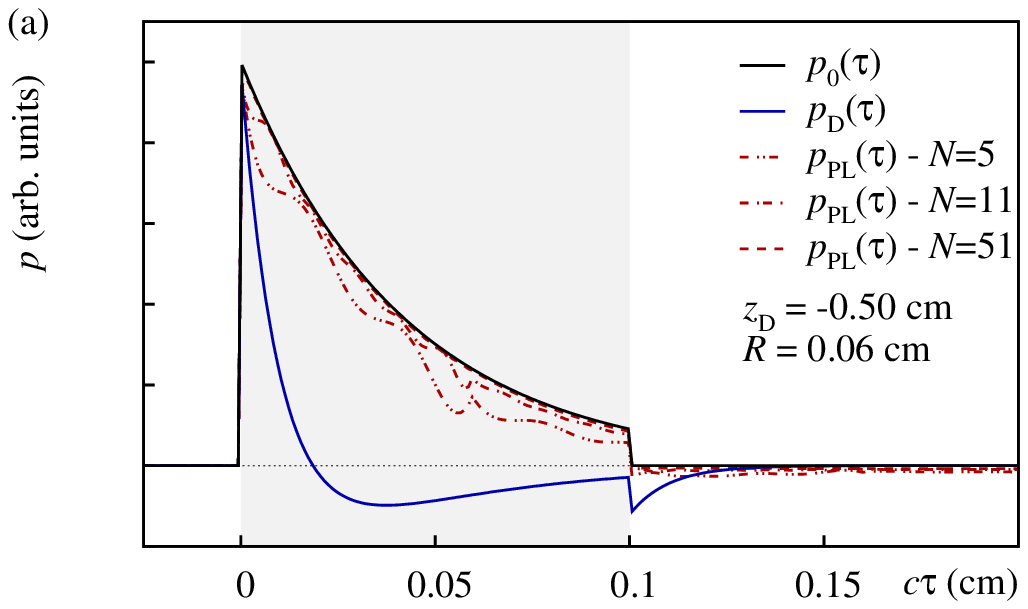}
\includegraphics[width=1.0\linewidth]{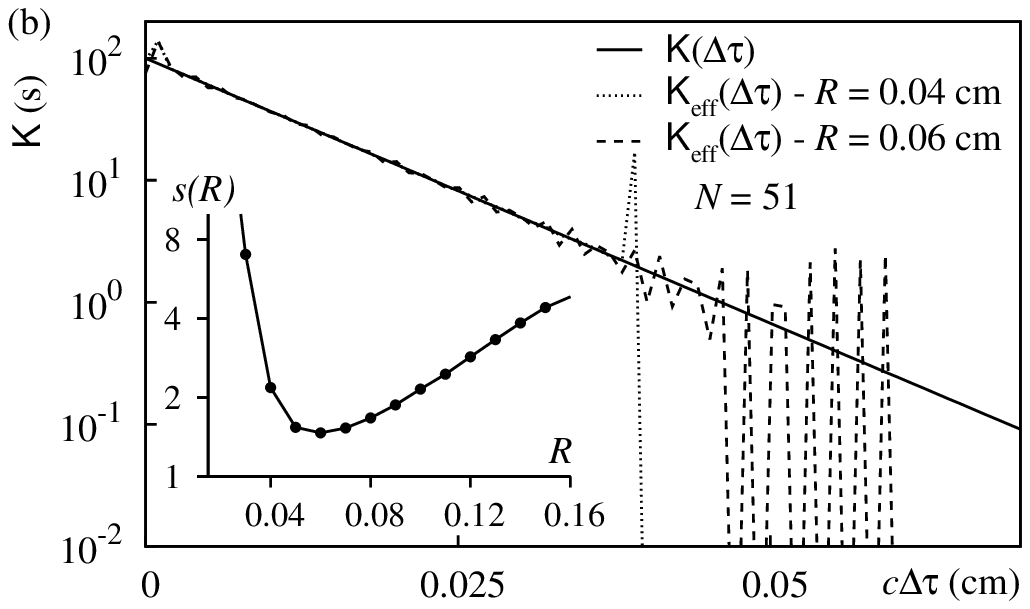}
\includegraphics[width=1.0\linewidth]{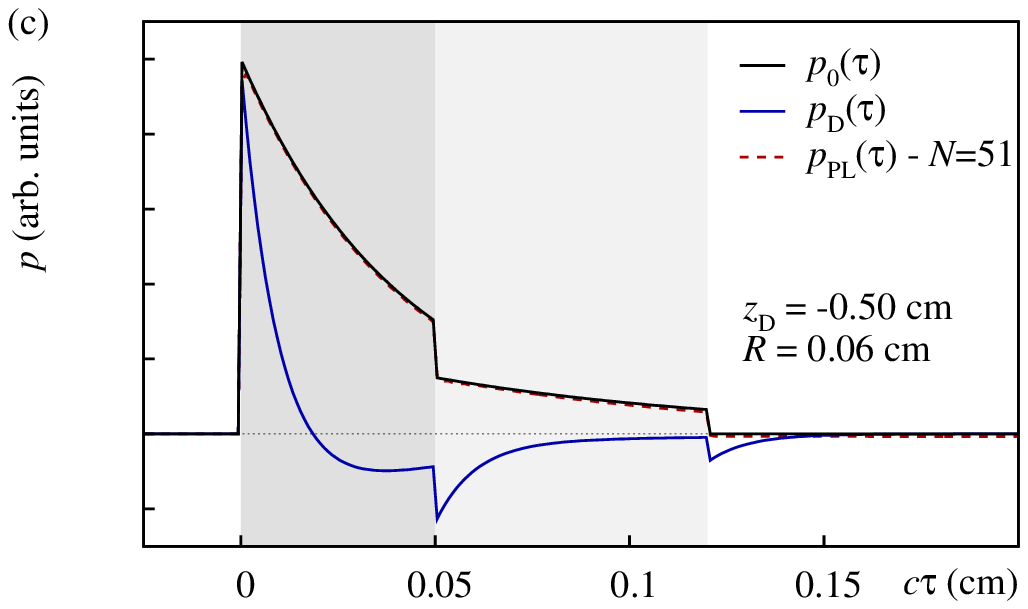}
\end{center}
\caption{(Color online) 
Kernel and source reconstruction within the paraxial approximation for system
parameters ${\bf p}_{\rm sys}=(c,a_{\rm B},z_{\rm D})\equiv(1\,{\rm cm/s},~0.1\,{\rm cm},~-0.5\,{\rm cm})$. 
(a) Inversion input $p_{\rm 0}$ (solid black line) and $p_{\rm D}$ (solid blue
line) used to derive effective kernel for $N=5$, $11$, and
$51$ Fourier-coefficients and cut-off parameter $R=0.06\,{\rm cm}$. Solution of
the respective source reconstruction problems yields the estimates $p_{\rm PL}$
(dashed and dash-dotted red curves).
(b) The main plot illustrates the effective kernel $\mathsf{K}_{\rm eff}(\Delta
\tau)\equiv \mathsf{K}(\Delta \tau; {\bf a}^{\star},R)$ for two different
cut-off distances $R=0.04\,{\rm cm}$, and $0.06\,{\rm cm}$. The inset shows the
SSR $s(R)\equiv s({\bf a}^\star,R)$ for $N=51$ as function of the cut-off
distance where the minimum is attained at $R=0.06\,{\rm cm}$.
(c) Solution $p_{\rm PL}$ of the source reconstruction problem for a 
OA signal $p_{\rm D}$ (solid blue line) resulting from a two-layer absorbing
structure for the same system
parameters as in (a). Source reconstruction is performed using the effective
kernel for ${\bf p}_{\rm rec}=(51,0.06\,{\rm
cm})$ resulting from the gauge procedure.
\label{fig:KRec}}
\end{figure}  

\paragraph{The inverse OA kernel reconstruction problem.}
Note that the solution of the direct problem and inverse problem (I.1) in terms
of Eq.\ (\ref{eq:OAVolterraInt}) is feasible using standard numerical schemes
based on, e.g., a trapezoidal approximation of the Volterra operator for a
generic kernel \cite{NumRec:1992}, or highly efficient memoization techniques
for the particular form of the above convolution kernel \cite{Stritzel:2016}.
As pointed out earlier, considering inverse problem (I.2), we here suggest a
Fourier-expansion of the Volterra kernel involving a sequence of $N$ expansion
coefficients ${\bf a}\equiv\{a_\ell\}_{0\leq \ell < N}$ and a cut-off distance
$R$ above which the resulting effective kernel is assumed to be zero, i.e.\
\begin{equation}
\mathsf{K}(x;{\bf a},R) = \sum_{\ell=0}^{N-1}a_\ell\,k_\ell(x; R)\,\Theta(R-x).
        \label{eq:KernelExpansion}
\end{equation}
The expansion functions $k_\ell(x;R)$ are given by 
\begin{align}
k_\ell(x;R) = 
 \begin{cases} 
 1,  & \text{if }\ell = 0 \\ 
        \cos\Big(2\pi\,\frac{\ell+1}{2} \frac{x}{R}\Big), 
        & \text{if } \ell ~\text{odd}\\
        \sin\Big(2\pi\,\frac{\ell}{2} \frac{x}{R}\Big), 
        & \text{if } \ell ~\text{even}
 \end{cases} 
 \label{eq:FourierFuncs}
\end{align}
and $\Theta(\cdot)$ signifies the Heavyside step-function.  Then, for a
suitable sequence ${\bf a}$, the Fourier approximation to
the Volterra integral equation, Eq.\ (\ref{eq:OAVolterraInt}), reads 
\begin{equation}
p_{\rm D}(\tau) = p_{\rm 0}(\tau) - 
        \sum_{\ell=0}^{N-1} a_\ell\,\mathsf{\Phi}_\ell(\tau;R), 
        \label{eq:FourierVolterraInt} 
\end{equation}
with reduced partial diffraction terms
\begin{equation}
\mathsf{\Phi}_\ell(\tau;R) = \int_{-\infty}^{\tau} \!k_\ell(\tau-\tau^\prime;R)\,
        \Theta(R-(\tau-\tau^\prime)) \,p_{\rm 0}(\tau^\prime)\,
        \mathrm{d}\tau^\prime.
        \label{eq:PhiFuncs}
\end{equation}
Now, consider a given set of input data $(p_{\rm 0},p_{\rm D})$ for known
apparative parameters ${\bf p}_{\rm sys}$, both in a discretized setting with
constant mesh interval $\Delta$, mesh points
$\{t_i\}_{0\leq i \leq M}$ where $t_0=0$, $t_i=t_{i-1}+\Delta$, and $t_M$ large
enough to ensure a reasonable measurement depth.  Then, bearing in mind that
$\tau_i=t_i+z_{\rm D}/c$, the optimal expansion coefficient sequence ${\bf
a}^\star$ can be obtained by minimizing the sum of the squared residuals (SSR)
\begin{equation}
s({\bf a},R) = \sum_{i=0}^{M}\Big[ (p_{\rm 0}(\tau_i)-p_{\rm D}(\tau_i)) 
        \,-\,\sum_{\ell=0}^{N-1} a_\ell \,\mathsf{\Phi}_\ell(\tau_i;R) \Big]^2.
        \label{eq:SSR}
\end{equation}
In the above optimization formulation of inverse problem (I.2), we considered a
trapezoidal rule to numerically evaluate the integrals that enter via the
functions $\mathsf{\Phi}_\ell(\tau_i;R)$.  In an attempt to construct an
effective Volterra kernel $\mathsf{K}(x;{\bf a},R)$ for a controlled setup with
{\emph{a priori}} known parameters ${\bf p}_{\rm sys}$, one might use the
high-precision ``Gaussian-beam'' estimator $a_\ell = (2 \omega_{\rm
D}/R)\int_{0}^{R}\!k_\ell(x;R)\,\exp\{-\omega_{\rm D} x\}\,\mathrm{d}x $ to
obtain an initial sequence ${\bf a}_{\rm 0}$ of expansion coefficients by means
of which a least-squares routine for the minimization of Eq.\ (\ref{eq:SSR})
might be started. In a situation where, say, $a_{\rm B}$ is only known
approximately or the assumption of a Gaussian beam profile is violated, one has
to rely on a rather low-precision coefficient estimate obtained by roughly
estimating the apparative parameters and resorting on the above
``Gaussian-beam'' estimate.

An exemplary kernel reconstruction procedure is shown in FIG.~\ref{fig:KRec},
where the OA signal $p_{\rm D}$ at ${\bf p}_{\rm sys}=(1\,{\rm cm/s},
~0.1\,{\rm cm},~-0.5\,{\rm cm})$, i.e.\ $D\approx3.75$, is first obtained by
solving the direct OA problem for Eq.\ (\ref{eq:OAVolterraInt}) for an
absorbing layer with $\mu_a=24\,{\rm cm^{-1}}$ in the range $z=0-0.1\,{\rm
cm}$, see black ($p_{\rm 0}$) and blue ($p_{\rm D}$) curves 
in FIG.~\ref{fig:KRec}(a). The set $(p_{\rm 0},p_{\rm D})$ is then used as
inversion input to compute the effective Volterra kernel for various sets of
reconstruction parameters ${\bf p}_{\rm rec}=(N,R)$. In particular, considering
$N=51$, the minimal value of $s({\bf a}^\star,R^\star)\approx 1.47$ is attained
at $R^\star=0.06\,{\rm cm}$, see the inset of FIG.~\ref{fig:KRec}(b).  As
evident from the main plot of FIG.~\ref{fig:KRec}(b), the effective Volterra
kernel for ${\bf p}_{\rm rec}=(51,R^\star)$ follows the exact stress wave
propagator for almost two orders of magnitude up to $c\Delta\tau\approx
0.05\,{\rm cm}$.  Beyond that limit, the noticeable deviation between both does
not seem to affect the overall SSR $s({\bf a},R)$ too much.  In this regard,
note that the kernel approximated for the (non optimal) choice ${\bf p}_{\rm
rec}=(51,0.04\,{\rm cm})$ exhibits a worse SSR. 

\begin{figure}[t!]
\begin{center}
\includegraphics[width=1.0\linewidth]{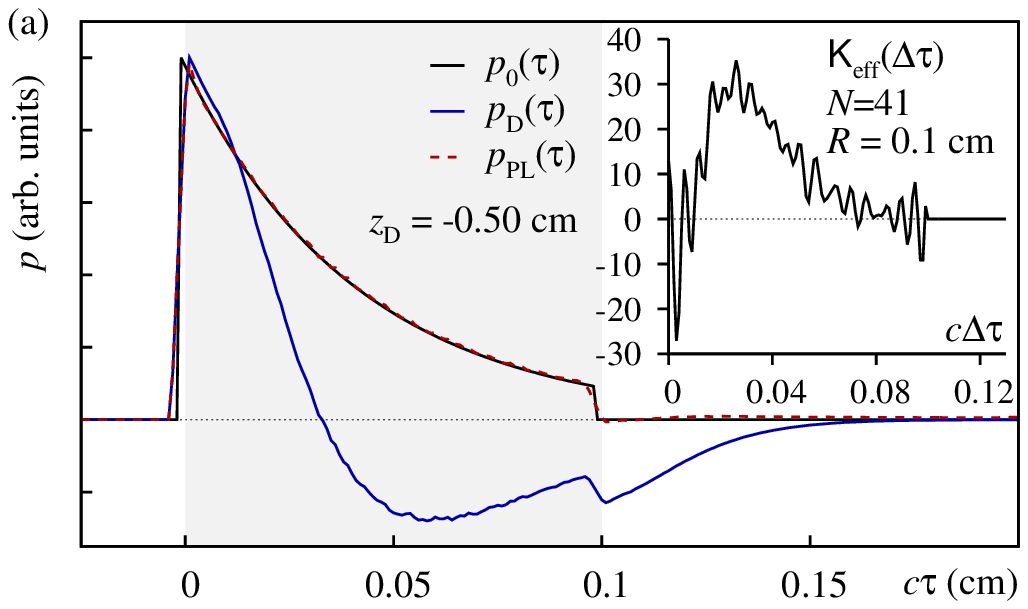}
\includegraphics[width=1.0\linewidth]{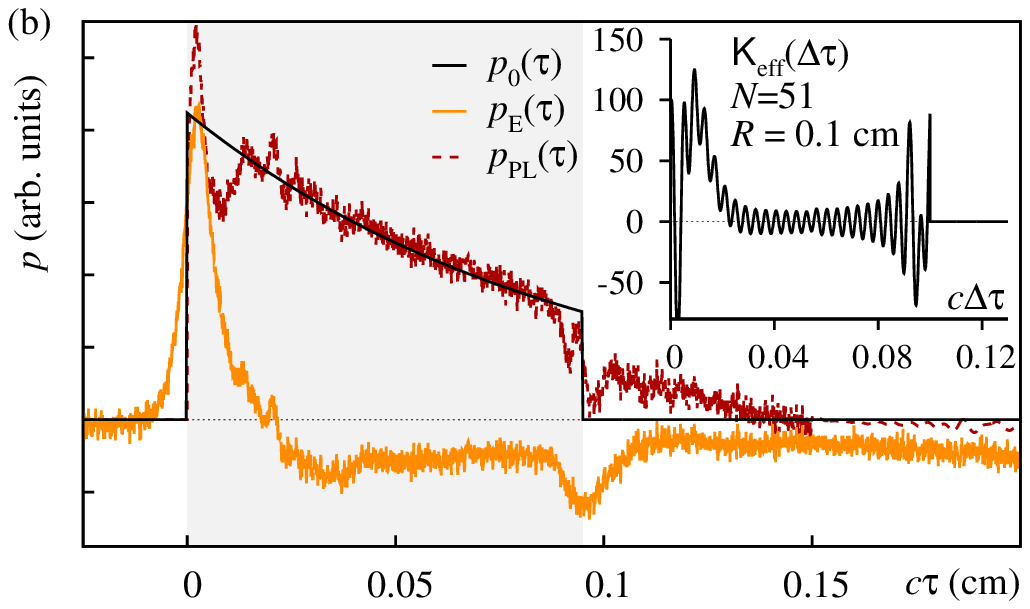}
\end{center}
\caption{(Color online) 
Inversion of OA signals to initial stress profiles beyond the paraxial 
approximation. Both figures illustrate the kernel and 
source reconstruction procedures for
(a) inversion of an OA signal featuring a top-hat irradiation source profile 
(see text). The main plot shows the input $(p_{\rm 0},p_{\rm D})$ to
the inversion procedure (solid black and blue lines, respectively) as well as
the reconstructed initial stress profile $p_{\rm PL}$ (dashed red line), and,  
(b) inversion of an OA signal resulting from an actual measurement
\cite{Blumenroether:2016}. The main plot shows the synthetic initial stress
profile $p_{\rm 0}$ (solid black line) used during the gauge procedure
as well as the inversion input $p_{\rm E}$ (orange line) for which the
reconstructed initial stress profile $p_{\rm PL}$ (dashed red line) is
obtained.
In both figures, the inset illustrates the effective Volterra kernel resulting
from the Fourier-approximation.
\label{fig:InvBeyondPA}}
\end{figure}  

\paragraph{The inverse OA source reconstruction problem.}
Note that the above Fourier-expansion approximation might be interpreted as a
gauge procedure to adjust an effective Volterra kernel $\mathsf{K}(x;{\bf
a}^\star,R)$ for an (possibly unknown) apparative setup ${\bf p}_{\rm sys}$,
here indirectly accessible through the diffraction transformation of the OA
signal $p_{\rm D}$ relative to $p_{\rm 0}$.  That is, once the kernel
reconstruction (I.2) is accomplished for a set of reference curves $(p_{\rm
0},p_{\rm D})_{\rm ref}$ under ${\bf p}_{\rm sys}$, the source reconstruction
problem (I.1) might subsequently be tackled also for all other OA signals
measured under ${\bf p}_{\rm sys}$ by solving the OA Volterra integral equation
Eq.\ (\ref{eq:OAVolterraInt}) in terms of a Picard-Lindel\"of ``correction''
scheme \cite{Hairer:1993}.  The latter
is based on the continued refinement of a putative solution, starting off from
a properly guessed ``predictor'' $p_{\rm PL}^{(0)}(\tau)$, improved
successively by solving 
\begin{equation}
p_{\rm PL}^{(n+1)}(\tau) = p_{\rm D}(\tau) + \int_{-\infty}^\tau 
        \!\mathsf{K}(\tau-\tau^\prime;{\bf a}^\star,R)\, 
        p_{\rm PL}^{(n)}(\tau^\prime)\,\mathrm{d}\tau^\prime.
        \label{eq:PicardIter}
\end{equation}
From a practical point of view we terminated the iterative correction scheme as
soon as the ${\rm max}$-norm $c_n\equiv \|p_{\rm PL}^{(n+1)}(\tau)-p_{\rm
PL}^{(n)}(\tau)\|$ of two successive solutions decreases below $c_n\leq
10^{-6}$. We here refer to the final estimate simply as $p_{\rm PL}$. Note
that, attempting a solution of (I.1) in the acoustic near-field, a
high-precision predictor can be obtained by using the initial guess $p_{\rm
PL}^{(0)}\equiv p_{\rm D}$. This is a reasonable choice since one might expect
the change of the OA near-field signal due to diffraction to be still quite
small. Further, source reconstruction in the acoustic far-field might be
started using a high-precision predictor obtained by integrating the OA signal
$p_{\rm D}$ in the far-field approximation \cite{Stritzel:2016}.
In contrast to this, low-precision predictors for both cases can be obtained by
setting $p_{\rm PL}^{(0)} \equiv c_{\rm 0}$, where, e.g., $c_{\rm 0}=0$.

The solution of the source reconstruction problem for the OA signal $p_{\rm D}$
used in the approximation of the Volterra kernel for the above setting ${\bf
p}_{\rm sys}=(1\,{\rm cm/s},\,0.1\,{\rm cm},\,-0.5\,{\rm cm})$ is shown in
FIG.~\ref{fig:KRec}(a). The apparent agreement of the data curves $p_{\rm PL}$
for ${\bf p}_{\rm rec}=(51,R^\star)$ and $p_{\rm 0}$ does not come as a
surprise since $p_{\rm D}$ was used for the gauge procedure in the first place.
As a remedy we attempt a source reconstruction for a second independent OA
signal, simulated for the same apparative setting only with two absorbing
layers $\mu_{a,1}=24\,{\rm cm^{-1}}$ from $z=0-0.05\,{\rm cm}$ and
$\mu_{a,2}=12\,{\rm cm^{-1}}$ from $z=0.05-0.12\,{\rm cm}$. As evident from
FIG.~\ref{fig:KRec}(c), inversion using the effective Volterra kernel from the
previous gauge procedure yields a reconstructed stress profile $p_{\rm PL}$ in
excellent agreement with the underlying exact initial stress profile $p_{\rm
0}$.

\paragraph{Inversion beyond the paraxial approximation.}
Given the apparent feasibility of the kernel reconstruction routine as a gauge
procedure to model the diffraction transformation of OA signals in terms of an
effective stress wave propagator in the framework of the OA Volterra integral
equation, we next address the inversion of OA signals to initial stress
profiles beyond the paraxial approximation.
Therefore, we first consider a borderline far-field signal for a top-hat
irradiation source 
\begin{equation}
f(\vec{r}_{\perp}) = 
\begin{cases} 
  1,    
        & \text{if } |\vec{r}_\perp| \leq \rho_0 \\ 
  \exp\{-(|\vec{r}_\perp| - \rho_0)^2/a_{\rm B}^2\},     
        & \text{if } |\vec{r}_\perp| > \rho_0
\end{cases}, 
        \label{eq:topHatISP}
\end{equation}
recorded at the system parameters ${\bf p}_{\rm sys}=(c,\rho_0,a_{\rm B},z_{\rm
D})=(1\,{\rm cm/s},\,0.1\,{\rm cm},\,0.1\,{\rm cm},\,-0.50\,{\rm cm})$, and
thus $D = 2|z_{\rm D}|/(\mu_{\rm a} (a_{\rm B}+\rho_0))\approx 1.04$, obtained
via an independent forward solver for the full OA wave equation 
designed for the solution of the
OA Poisson integral for layered media \cite{Wang:2009,Blumenroether:2016}. The
inversion results are summarized in FIG.~\ref{fig:InvBeyondPA}(a), where the
kernel reconstruction (inset) and source reconstruction (main plot) are shown
for the parameter set ${\bf p}_{\rm rec}=(41,\,0.1\,{\rm cm})$.  The excellent
agreement of the stress profiles $p_{\rm 0}$ and $p_{\rm PL}$ suggests that the
kernel reconstruction routine also applies to a more general OA setting, based
on the full OA wave equation. 
Finally, we consider an OA signal resulting from an actual measurement on PVA
hydrogel based tissue phantoms \cite{Blumenroether:2016}. In this case we
carefully estimated the apparative parameters ${\bf p}_{\rm sys}=(150000\,{\rm
cm/s},\,0.054\,{\rm cm},0.081{\rm cm/s},\,-0.3\,{\rm cm})$ as well as
$\mu_{a}=11\,{\rm cm}$ in the range $z=0-0.095\,{\rm cm}$, i.e.\
$D\approx6.73$, in order to create a set of synthetic input data by means of
which an appropriate kernel gauge procedure can be carried out.  The result of
the procedure using ${\bf p}_{\rm rec}=(51,\,0.1\,{\rm cm})$ is shown in
FIG.~\ref{fig:InvBeyondPA}(b). So as to perform the source reconstruction for
the experimental signal $p_{\rm E}$, we considered data within the interval
$c\tau = [0,\,0.15]\,{\rm cm}$, only. As evident from the figure, the
reconstructed stress profile $p_{\rm PL}$ fits the signal $p_{\rm 0}$ used in
the gauge procedure remarkably well \footnote{A {\sc Python} implementation of our
code for the solution of inverse problems (I.1) and (I.2) can be found at
\url{https://github.com/omelchert/INVERT.git}.}.

\paragraph{Conclusions.}
In the presented Letter we have introduced and discussed the kernel
reconstruction problem in the paraxial approximation to the optoacoustic wave
equation. We suggested a Fourier-expansion approach to approximate the Volterra
kernel which takes a central role in the theoretical framework.  The developed
approach proved useful as gauge procedure by means of which the diffraction
transformation experienced by OA signals can effectively be modeled, allowing
to subsequently solve the source reconstruction problem in the underlying
apparative setting.  From this numerical study we found that the developed
approach extends beyond the framework of the paraxial approximation and also
allows for the inversion of OA signals described by the full OA wave equation.
From a point of view of computational theoretical physics it would be tempting
to explore other kernel expansions in terms of generalized Fourier series as
well as gauge procedures involving sets of measured pressure profiles only.
Such investigations are currently in progress with the aim to shed some more
light on this intriguing inverse problem in the field of optoacoustics and to
facilitate a complementary approach to conventional OA imaging.

\paragraph{Acknowledgments.}
We thank A.~Demircan for commenting on an early draft of the manuscript and
E.~Blumenr\"other for providing experimental data.  This research work received
funding from the VolkswagenStiftung within the ``Nieders\"achsisches Vorab''
program in the framework of the project ``Hybrid Numerical Optics''  (HYMNOS;
Grant ZN 3061).  Valuable discussions within the collaboration of projects
MeDiOO and HYMNOS at HOT are gratefully acknowledged.

\bibliography{masterBibfile_optoacoustics,commentsBibfile_optoacoustics}

\end{document}